\documentstyle[12pt]{article}
\font\tenrm=cmr10
 at 10truept

\def\fr#1#2{{{#1} \over {#2}}}

\def\frac#1#2{{\textstyle{{#1}\over {#2}}}}

\def\lsim{\mathrel{\rlap{\lower4pt\hbox{\hskip1pt$\sim$}}
    \raise1pt\hbox{$<$}}}
\def\gsim{\mathrel{\rlap{\lower4pt\hbox{\hskip1pt$\sim$}}
    \raise1pt\hbox{$>$}}}
\def\sqr#1#2{{\vcenter{\vbox{\hrule height.#2pt
         \hbox{\vrule width.#2pt height#1pt \kern#1pt
         \vrule width.#2pt}
         \hrule height.#2pt}}}}

\newcommand{\beq}{\begin{equation}}
\newcommand{\eeq}{\end{equation}}
\newcommand{\bea}{\begin{eqnarray}}
\newcommand{\eea}{\end{eqnarray}}

\begin{document}
\titlepage

\begin{flushright}
{IUHET 305\\}
{COLBY 95-03\\}
{June 1995\\}
\end{flushright}
\vglue 1cm

\begin{center}
{{\bf THE REVIVAL STRUCTURE OF RYDBERG WAVE PACKETS\\
   BEYOND THE REVIVAL TIME\footnote[1]{\tenrm
Paper presented by R.B. at the
Seventh Rochester Conference on Coherence
and Quantum Optics,
Rochester, NY, June 1995
}
\\}
\vglue 1.0cm
{Robert Bluhm$^a$ and V. Alan Kosteleck\'y$^b$\\}
\bigskip
{\it $^a$Physics Department, Colby College\\}
\medskip
{\it Waterville, ME 04901, U.S.A.\\}
\vglue 0.3cm
{\it $^b$Physics Department, Indiana University\\}
\medskip
{\it Bloomington, IN 47405, U.S.A.\\}
\vglue 0.3cm
\bigskip

\vglue 0.8cm
}
\vglue 0.3cm

\end{center}

{\rightskip=3pc\leftskip=3pc\noindent
After a Rydberg wave packet forms,
it is known to undergo a series of collapses
and revivals within a time period called the revival time
$t_{\rm rev}$,
at the end of which it resembles its original shape.
We study the behavior of Rydberg wave packets
on time scales much greater than
$t_{\rm rev}$.
We find that after a few revival cycles the wave packet
ceases to reform at multiples of the revival time.
Instead,
a new series of collapses and revivals commences,
culminating after a time period $t_{\rm sr} \gg t_{\rm rev}$
with the formation of a wave packet that more closely
resembles the initial packet than does the full
revival at time $t_{\rm rev}$.
Furthermore,
at times that are rational fractions of $t_{\rm sr}$,
we show that the motion of the wave packet is periodic
with periodicities that can be expressed
as fractions of the revival time $t_{\rm rev}$.
These periodicities indicate a new type of fractional revival,
occurring for times much greater than $t_{\rm rev}$.
We also examine the effects of quantum defects and
laser detunings on the revival structure of Rydberg wave
packets for alkali-metal atoms.

}

\vskip 0.2truein
\centerline{\it
To appear in }
\centerline{\it
Coherence and Quantum Optics VII}
\centerline{\it
J. Eberly, L. Mandel, and E. Wolf, editors,}
\centerline{\it
Plenum, New York, 1996}

\vfill
\newpage

\begin{center}
\pagestyle{empty}
\baselineskip=14pt
{{\bf THE REVIVAL STRUCTURE OF RYDBERG WAVE PACKETS\\
   BEYOND THE REVIVAL TIME\footnote[1]{\tenrm
Paper presented by R.B.}\\}
\vglue 1.0cm
{Robert Bluhm$^a$ and
V. Alan Kosteleck\'y$^b$
\\}
\bigskip
{\it $^a$Physics Department, Colby College\\}
\medskip
{\it Waterville, ME 04901, U.S.A.\\}
\vglue 0.3cm
{\it $^b$Physics Department, Indiana University\\}
\medskip
{\it Bloomington, IN 47405, U.S.A.\\}}
\end{center}

\vspace{0.1in}
\pagestyle{empty}

When a Rydberg atom is excited by a short laser pulse,
a state is created that has classical behavior for a
limited time
[1].
The wave packet initially oscillates with the classical
keplerian period $T_{\rm cl} = 2 \pi {\bar n}^3$,
where ${\bar n}$ is the mean value of the
principal quantum number excited in the packet.
However,
the motion is not entirely classical
because the wave packet disperses with time.
After many Kepler orbits,
at the revival time $t_{\rm rev}$
the wave packet recombines nearly into its original shape.
Moreover,
prior to this full revival,
the wave function evolves through a sequence of
fractional revivals.
Experiments have detected all stages of the evolution
of the wave packet during the time $t_{\rm rev}$,
including the initial classical motion,
the full revival, and the fractional revivals
[2].

In this paper,
we summarize results concerning
the time evolution and revival structure
of Rydberg wave packets
on time scales much greater than the revival time $t_{\rm rev}$.
We have found that a new system of full and fractional revivals
occurs for times beyond the revival time,
with structure different from that of the usual fractional revivals.
We have also examined the effects of quantum defects and
laser detunings on the revival structure of Rydberg wave
packets for alkali-metal atoms.
The results summarized here have been published in
refs.\ [3,4].

The time-dependent wave function for a Rydberg wave packet
may be written as an expansion in terms of hydrogenic
energy eigenstates:
\beq
\Psi ({\vec r},t) = \sum_{n} c_n
\varphi_n({\vec r}) \exp \left( -i E_n t \right)
\quad ,
\label{psi}
\eeq
where $\varphi_n ({\vec r})$ is a hydrogenic wave function
and $c_n = \left< \Psi (0) \vert \varphi_n \right>$
is a weighting coefficient.
We expand $E_n$ around $\bar n$:
\beq
E_n \simeq E_{\bar n} + E_{\bar n}^\prime (n - {\bar n})
+ \fr 1 2 E_{\bar n}^{\prime\prime} (n - {\bar n})^2
+ \fr 1 6 E_{\bar n}^{\prime\prime\prime} (n - {\bar n})^3
+ \cdots
\quad ,
\label{energy}
\eeq
where each prime on $E_{\bar n}$
denotes a derivative.
This expansion defines three distinct time scales:
$T_{\rm cl} = \fr {2 \pi} {E_{\bar n}^\prime} = 2 \pi {\bar n}^3$,
$t_{\rm rev} = \fr {- 2 \pi} {\fr 1 2 E_{\bar n}^{\prime\prime}}
= \fr {2 {\bar n}} 3 T_{\rm cl}$,
and
$t_{\rm sr} = \fr {2 \pi} {\fr 1 6 E_{\bar n}^{\prime\prime\prime}}
= \fr {3 {\bar n}} 4 t_{\rm rev}$.
These time scales
determine the evolution and
revival structure of the wave packet.
The first two are the usual time scales
relevant in the description of the conventional revival structure.
We include the third-order term
because we are interested in times much greater
than the revival time.
This term defines the new time scale,
which we refer to as the superrevival time
$t_{\rm sr}$.

By examining the effects of all three terms in
the expansion of the wave function,
we have found that at certain times $t_{\rm frac}$
it is possible to expand
the wave function $\Psi ({\vec r},t)$
as a series of subsidiary wave functions.
We find that when
$t_{\rm frac} \approx \fr 1 q t_{\rm sr}$,
where $q$ is an integer multiple of 3,
the wave packet can be written as a sum of
macroscopically distinct wave packets.
Furthermore,
at these times $t_{\rm frac}$
we also find that the motion of the wave packet is
periodic with a period
$T_{\rm frac} \approx \fr 3 q t_{\rm rev}$.
Note that these periodicities are different
from those of the fractional revivals,
and thus a new level of revivals commences
for $t > t_{\rm rev}$.
We also find that at the particular time
$t_{\rm frac} \approx \fr 1 6 t_{\rm sr}$,
a single wave packet forms that resembles
the initial wave packet more closely than the
full revival does at time $t_{\rm rev}$,
i.e., a superrevival occurs.

These results can be generalized to include the effects
of quantum defects,
corresponding to nonhydrogenic energies $E_{n^\ast}$,
and also to the case where
the laser excites a mean energy corresponding to
a noninteger value $N^\ast$.
We find that the effects of quantum defects
on the occurrence times and
periodicities of long-term revivals are different from
those of the laser detunings.
A laser detuning cannot be mimicked by quantum defects
or vice versa.
Furthermore,
the modification to the long-term revival
times induced by the quantum defects cannot be obtained by
direct rescaling of the hydrogenic results.

It is feasible that
an experiment can be performed to detect
the full and fractional revival structure
for $t \gg t_{\rm rev}$
discussed in this paper.
One possibility is
to use the pump-probe time-delayed photoionization
method of detection
for radial Rydberg wave packets excited in
alkali-metal atoms with
${\bar n} \approx 45$ -- $50$,
provided a delay line of 3 -- 4 nsec
is installed in the apparatus.
For smaller values of $\bar n$,
the required delay times can be reduced below 1 nsec.
With ${\bar n} \simeq 36$,
for example,
the full/fractional superrevivals could be detected
with delay lines used currently in experiments.

\vglue 0.1cm

This work is supported in part by the National
Science Foundation under grant number PHY-9503756.

\vglue 0.1cm

\begin{enumerate}
\item
J. Parker and C.R.\ Stroud,
Phys.\ Rev.\ Lett.\ {56}, 716 (1986);
G.\ Alber, H.\ Ritsch, and P.\ Zoller,
Phys.\ Rev.\ A {34}, 1058 (1986);
I.Sh.\ Averbukh and N.F.\ Perelman,
Phys.\ Lett.\ {139A}, 449 (1989);
M. Nauenberg,
J. Phys.\ B {23}, L385 (1990).
\item
For additional references,
see [1]
and [3].

\item
R. Bluhm and V.A. Kosteleck\'y,
Phys.\ Lett.\ { 200A}, 308 (1995)
quant-ph/9508024;
R. Bluhm and V.A. Kosteleck\'y,
Phys.\ Rev.\ A { 51}, 4767 (1995)
quant-ph/9506009.

\item
R. Bluhm and V.A. Kosteleck\'y,
Phys.\ Rev.\ A { 50}, R4445 (1994)
hep-ph/9410325.

\end{enumerate}

\vfill\eject

\end{document}